\documentstyle[nato,psfig]{crckapb} 

\newcommand{\Msun}{M_{\odot}}
\newcommand{\CO}{^{12}\mathrm{C}(\alpha,\gamma)^{16}\mathrm{O}}

\begin{opening}
\title{Stellar yields and chemical evolution}
\subtitle{The solar neighborhood as a calibrator}

\author{Daniel Thomas}
\institute{Universit\"ats-Sternwarte M\"unchen}
\author{Laura Greggio}
\institute{Osservatorio Astronomico di Bologna}
\author{Ralf Bender}
\institute{Universit\"ats-Sternwarte M\"unchen}
\end{opening}

\runningtitle{Stellar yields and chemical evolution}

\begin{document}

\begin{abstract}
Uncertainties in stellar nucleosynthesis and their impact on models of
chemical evolution are discussed. Comparing the Type II supernova
nucleosynthesis prescriptions from Woosley \& Weaver (1995) and Thielemann,
Nomoto, \& Hashimoto (1996), it turns out that the latter predict higher Mg/Fe
ratios that are more favorable in reproducing the observed abundance features
of the Milky Way. Provided that chemical evolution models are calibrated on
the solar neighborhood, they offer a powerful tool to constrain structure
formation. In particular, galaxy formation models that yield star formation
histories significantly longer than 1~Gyr fail to reproduce the super-solar
Mg/Fe ratios observed in elliptical galaxies.
\end{abstract}

\section{Introduction}
Population synthesis models based on {\em solar} abundance ratios
predict -- for a given Fe index -- Mg indices that are weaker than measured in
the integrated spectra of elliptical galaxies (e.g.\ Worthey, Faber \&
Gonz\'{a}lez 1992 \cite{WFG92}; Davies, Sadler \& Peletier 1993 \cite{DSP93};
Mehlert et al.\ 2000 \cite{Mehetal00}). Although the link from line indices to
element abundance ratios is not straightforward (Greggio 1997 \cite{Gre97};
Tantalo, Chiosi, \& Bressan 1998 \cite{TCB98}), the strong Mg absorption
features are interpreted as an enhancement of $\alpha$-elements. More
quantitative studies find an average [Mg/Fe] overabundance of $0.3-0.4$ dex
(Weiss, Peletier, \& Matteucci 1995 \cite{WPM95}; Greggio 1997
\cite{Gre97}). This conclusion gets further support from the detection of
enhancement of Mg and other $\alpha$-elements in the stars of the Milky Way
bulge (McWilliam \& Rich 1994 \cite{MW94}). These findings imply short
formation timescales and/or an initial mass function (IMF) that is flat at the
high-mass end (Matteucci 1994 \cite{Ma94}). Scenarios that form elliptical
galaxies in mergers of evolved spirals fail to reproduce such
$\alpha$-enhanced stellar populations unless a significant flattening of the
IMF is assumed (Thomas, Greggio, \& Bender 1999 \cite{TGB99}). Galaxy
formation models that are based on hierarchical clustering lead to extended
star formation histories in elliptical galaxies which produces Mg/Fe element
ratios that are too low compared to the values quoted above (Bender 1996
\cite{Be96}; Thomas 1999 \cite{Th99}; Thomas \& Kauffmann 2000 \cite{TK00}).

In this paper we address the question of how robust the predictions from
chemical evolution models are. For this purpose we analyze two recent Type II
supernova (SNII) nucleosynthesis prescriptions (Woosley \& Weaver 1995
\cite{WW95}, hereafter WW95; Thielemann, Nomoto, \& Hashimoto 1996
\cite{TNH96} and Nomoto et al.\ 1997 \cite{Netal97}, hereafter TNH96), that
represent a crucial input for the determination of Mg/Fe element ratios as a
function of timescales (see also Gibson 1997 \cite{G97b}; Portinari, Chiosi,
\& Bressan 1998 \cite{PCB98}).

\section{Stellar yields}
The ratio of Mg to Fe is a measure for the contribution from high-mass stars
to the chemical enrichment, because Mg is only produced in SNII (short-lived,
massive stars), while a significant part of Fe comes from Type Ia supernovae
(SNIa, low-mass, long-lived binary systems). The crucial input ingredients to
calculate the enrichment of these elements are both the supernova rates and
the stellar yields (nucleosynthesis and ejection per supernova, see Chieffi,
this volume; Limongi, this volume). In the following we briefly comment
on the SNII yields of Mg and Fe that are calculated by WW95 and TNH96
(see also Thomas, Greggio, \& Bender 1998 \cite{TGB98}).

\paragraph{Magnesium}
Magnesium is produced in both hydrostatic (before the explosion) and explosive
carbon burning. The synthesis thus depends upon the (still uncertain)
parameters of stellar evolution (convection criteria, $\CO$-rate, etc.) and
the modeling of the explosion. We find that the yields of Mg in the two sets
of models are in good agreement for low initial stellar masses in the range
$10-18~\Msun$. In a 20~$\Msun$ star, instead, TNH96 produce significantly more
Mg than WW95, likely because they adopt the Schwarzschild criterion for
convection (WW95; Thomas et al.\ 1998 \cite{TGB98}). At the high-mass end
($30-70~\Msun$), the total mass and the composition of the ejecta mainly
depend on the explosion energy and on the mass-cut, which leads to slightly
higher Mg-yields in the TNH96 models.

\paragraph{Iron}
Since the iron {\em ejected} by SNeII is entirely produced during the
explosion, the yield is very sensitive to details of the explosion model
(total energy, mass-cut, fall-back effect, energy transport, etc.). Hence,
particularly at higher masses, the Fe-yields from SNeII are very uncertain.
For $M<35~\Msun$, WW95 (model B) give higher Fe yields than TNH96.

\begin{figure}[ht]
\begin{center}
\psfig{figure=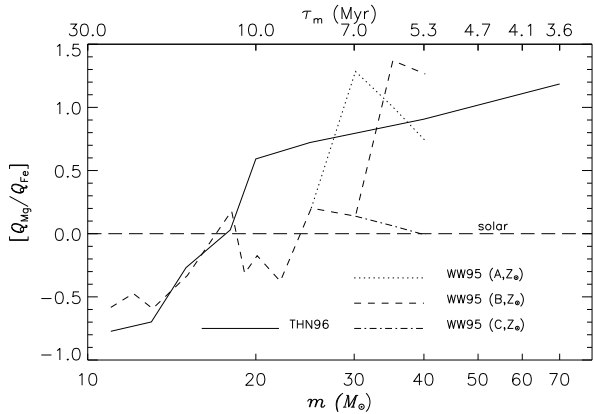,width=0.65\textwidth}
\end{center}
\caption{[Mg/Fe] ratio in the ejecta of SNII explosions as a function
of initial stellar mass. WW95 with solar initial metallicity are considered.
WW95 models A,B,C assume different explosion energies for the high-mass stars
increasing from model A to C. Model B is equivalent to the prescription in
TNH96.}
\label{thomas1}
\end{figure}
Fig.~\ref{thomas1} shows the resulting [Mg/Fe] ratio in the SNII ejecta for
the two nucleosynthesis prescriptions WW95 and TNH96. For $M\sim 20~\Msun$,
the [Mg/Fe] predicted by TNH96 is $\sim 1$~dex higher than WW95. At the
high-mass end, WW95 (model B) produce larger Mg/Fe because of the very low Fe
yield due to the fall-back effect.

\section{The solar neighborhood}
The significant discrepancies between the Mg/Fe yields from WW95 and TNH96
lead to very different conclusions on star formation timescales and IMF slopes
(Thomas et al.\ 1998 \cite{TGB98}, 1999 \cite{TGB99}). It is therefore crucial
to calibrate the chemical evolution model on the abundance patterns of the
solar neighborhood, for which the observational data provide the most detailed
information. Here we show the impact of the above discrepancies in the stellar
yields on the evolution of [Mg/Fe] with [Fe/H] in the solar neighborhood. The
model description for the chemical evolution in the Milky Way is based on the
standard infall model (Matteucci \& Greggio 1986 \cite{MG86}). The recipe for
the rate of SNIa is taken from Greggio \& Renzini (1983) \cite{GR83}, a
Salpeter initial mass function ($x=1.35$, $M_{\rm min}=0.1~\Msun$, $M_{\rm
max}=70~\Msun$) is adopted. For more details see Thomas et al.\ (1998)
\cite{TGB98}.

\begin{figure}[ht]
\psfig{figure=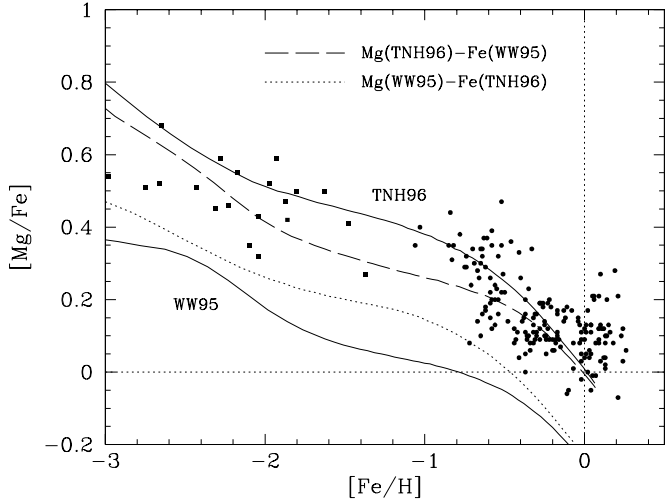,width=0.75\textwidth}
\caption{[Mg/Fe] ratio as a function of iron abundance [Fe/H] in the solar
neighborhood. Data are from Magain 1989 (squares) and Edvardsson et al.\ 1993
(circles). The solid lines denote models using SNII-yields from WW95 and
TNH96. The broken lines show the case for mixed stellar yields.
Mg(TNH96)-Fe(WW95): dashed; Mg(WW95)-Fe(TNH96): dotted.}
\label{thomas2}
\nocite{Ma89,EAGLNT93}
\end{figure}
Fig.~\ref{thomas2} shows observed stellar abundance ratios in the
[Mg/Fe]-[Fe/H] plane. Models using the same input parameters but different
sets of SNII nucleosynthesis prescriptions (WW95 and TNH96) are plotted as
solid lines. With the TNH96-yields, the Mg-enhancement in the (metal-poor)
halo stars (${\rm [Fe/H]}<-1$) is well reproduced. The sharp decrease of
[Mg/Fe] at ${\rm [Fe/H]}\sim -1$ implies the enrichment from SNIa to be
delayed by $\sim 1$~Gyr (Pagel \& Tautvaisiene 1995 \cite{PT95}, this volume)
which is well matched with the SNIa model adopted here (see also Greggio 1996
\cite{Gr96}). Adopting WW95 yields without modifying the other input
parameters, the resulting Mg/Fe ratios are well below the data for all
metallicities (see also Timmes, Woosley, \& Weaver 1995
\cite{TWW95}). Note that the strongest constraint on the SNII nucleosynthesis
comes from the metal-poor halo stars, while the mismatch around solar
metallicity could in principle be improved by adjusting the parameters of the
star formation history and of the SNIa rate.

Models with mixed stellar yields, namely, Mg(TNH96)-Fe(WW95) and
Mg(WW95)-Fe(TNH96) are shown by the dashed and dotted lines, respectively.
This exercise demonstrates that the low Mg/Fe ratios deduced from the WW95
models is mainly due to the low Mg yields. Especially at the high metallicity
regime, the large Fe SNII-yields from WW95 have only a small effect on the
final abundance ratios as a considerable fraction ($\sim 60$ per cent) of Fe
comes from SNeIa.

We conclude that the impact of uncertainties in stellar yields on chemical
evolution models is not negligible. Chemical evolution models that are applied
to the Bulge or extragalactic systems should be carefully calibrated on the
solar neighborhood.

\section{Discussion}
\subsection{Model improvements}
An important requirement for the successful calibration of chemical evolution
models is the accuracy and reliability of abundance and age determinations.
Analyzing high quality spectra of galactic stars, Fuhrmann (1998) \cite{Fu98}
demonstrates that the scatter in Mg/Fe found by Edvardsson et al.\
(1993) \cite{EAGLNT93} can be reduced. His data also point towards a more
pronounced separation of the various components of the Galaxy, that may
require more sophisticated modeling (e.g., two-infall model, Chiappini,
Matteucci, \& Gratton 1997 \cite{CMG97}). 

The determination of the exact timescale for the delayed Fe enrichment from
SNIa depends on the star formation history that is constrained through the
observationally defined age-metallicity relation. As the latter still suffers
from a large scatter, it may be more promising in the future to directly adopt
the star formation history that is determined with chromospheric ages
(Rocha-Pinto et al.\ 2000 \cite{Retal00}).

Invoking a metallicity dependence of SNIa rates (Kobayashi et al.\ 1998
\cite{Ketal98}) and the consideration of delayed and inhomogeneous mixing
(Malinie et al.\ 1993 \cite{Metal93}; Thomas et al.\ 1998 \cite{TGB98};
Tsujimoto, Shigeyama, \& Yoshii 1999 \cite{TSY99}) represent further
improvements of chemical evolution models.

\subsection{Constraint on elliptical galaxies}
Unless one allows for a flattening of the IMF, the very high Mg/Fe ratios in
elliptical galaxies prohibit the occurrence of late star formation, i.e.\
induced by a merger (Thomas et al.\ 1999 \cite{TGB99}). This constraint seems
to be at odds with the finding that ellipticals exhibit a considerable scatter
in H$\beta$ line strengths (Gonz\'alez 1993 \cite{G93}). However, composite
stellar population models containing a small fraction of old metal-poor stars
allow for an alternative interpretation: Such models succeed in reproducing
the strong metallic {\em and} the strong Balmer lines observed in ellipticals
without invoking young ages (Maraston, this volume; Maraston \& Thomas 2000
\cite{MT00}).

\section*{Acknowledgments}
DT thanks F.\ Matteucci and F.\ Giovannelli for a very lively and fruitful
workshop. This work was supported by the "Sonderforschungsbereich 375-95 f\"ur
Astro-Teilchenphysik" of the Deutsche Forschungsgemeinschaft.


\end{document}